\documentstyle[aps,multicol]{revtex}
\begin{document}

\newcommand{\av}[2]{\left\langle #1 \right\rangle_{#2}}
\newcommand{\ep}{\varepsilon}
\newcommand{\om}{\omega}
\newcommand{\imag}{{\rm Im}\;}
\newcommand{\real}{{\rm Re}\;}
\newcommand{\tr}{{\rm tr}\;}
\def\BRA{\left\langle}
\def\KET{\right\rangle}
\def\BBRA{\BRA\BRA}
\def\KKET{\KET\KET}
\def\LMID{\left.\mid}
\def\RMID{\mid\right.}
\def\LK{\left(}
\def\RK{\right)}
\def\LBK{\left\lbrack}
\def\RBK{\right\rbrack}
\def\LB{\left\lbrace}
\def\RB{\right\rbrace}
\def\a{\alpha}

\draft
\title{Perturbation Theory for the Rosenzweig-Porter Matrix Model}
\author{Alexander Altland$^{1,2}$, Martin Janssen$^1$ and Boris
Shapiro$^3$}
\address{$^1$ Institut f\"ur Theoretische Physik, Universit\"at
zu K\"oln, Z\"ulpicher Strasse 77, 50937 K\"oln, Germany \\\ $^2$
Cavendish Laboratory, Madingley Road, Cambridge CB3 OHE, UK\\\
$^3$ Department of Physics, Technion-Israel Institute of Technology, 32000 Haifa, Israel} 
\date{May 5, 1997} 
\maketitle

\begin{abstract}
We study an ensemble of random matrices (the Rosenzweig-Porter model)
which, in contrast to the standard Gaussian ensemble, is not invariant
under changes of basis.  We show that a rather complete understanding
of its level correlations can be obtained within the standard
framework of diagrammatic perturbation theory. The structure of the
perturbation expansion allows for an interpretation of the level
structure on simple physical grounds, an aspect that is missing in the
exact analysis (T. Guhr, Phys. Rev. Lett. {\bf 76}, 2258 (1996),
T. Guhr and A. M\"uller-Groeling, cond-mat/9702113).
\end{abstract}

\pacs{PACS number: 05.45.+b Theory and models of chaotic systems}

\begin{multicols}{2}
\section{Introduction}
Random matrix ensembles were introduced into physics by Wigner, Dyson
and others \cite{Meh91} as phenomenological models of complex quantum
systems. Such ensembles are defined as to obey certain symmetries and
constraints but are otherwise ``as random as possible''. For instance,
the Gaussian unitary ensemble (GUE) consists of all $N\times N$
Hermitian matrices $H$, the only constraint being that, on the
average, $\tr H^2$ is a given constant. This leads to a probability
density in the matrix space, ${\cal P}_0 \sim \exp \LK -\tr H^2 \RK$,
which is invariant under unitary transformations.

Recently there has been some interest in various generalizations of
the GUE and its orthogonal and symplectic counterparts \cite{Sha96}.
One possible generalization amounts to breaking the $U(N)$ symmetry of
the GUE by introducing a parameter $\mu$ into the probability density
function
\begin{eqnarray}\label{1.1}
        &&{\cal P} (H) dH = {\cal N}\, 
        \exp{(-F(H))}dH \, ,\\
        &&F(H)=  \sum_{i=1}^N
        H_{ii}^2 + 2\LK 1+\mu\RK \sum_{i<j} \LBK (\real  H_{ij})^2 +
        (\imag  H_{ij})^2\RBK .\nonumber
\end{eqnarray}
Here $H_{ij}$, with $i\leq j$, designate the independent matrix
elements of a $N\times N$ Hermitian matrix, $dH=\prod\limits_{i=1}^N
dH_{ii} \prod\limits_{i<j} d(\real H_{ij}) \, d(\imag H_{ij})$ is the
volume element in the matrix space and ${\cal N}$ is a normalization
constant.  For $\mu=0$ the expression in the curly brackets is equal
to $(-\tr H^2)$, so that the GUE is recovered. The parameter $\mu$
breaks the $U(N)$ symmetry and introduces a preferential basis. When
$\mu\to\infty$, for fixed $N$, all matrices become diagonal in that
basis. The ensemble, thus, exhibits a crossover from the Wigner-Dyson
statistics of the standard random matrix theory ($\mu=0$) to the
Poisson statistics of uncorrelated levels ($\mu=\infty$). Such an
ensemble (for real symmetric matrices) was introduced by Rosenzweig
and Porter \cite{Ros60} in their studies of complex atomic spectra,
and more recently appeared in the field of quantum chaos and
localization \cite{Sha96}.

We shall be interested in the behavior of this ensemble in the
$N\to\infty$ limit. In this limit, significant deviations from the
GUE-behavior can only occur if $\mu$ increases with $N$ sufficiently
fast.  The local level statistics is controlled
\cite{Jac95} --\cite{Wei96} by the parameter $\mu/N^2$. Only when this
parameter approaches infinity, the Wigner-Dyson statistics becomes
completely obliterated and the Poisson limit of uncorrelated levels is
reached. In the opposite case, i.e.~when $(\mu/N^2)\to 0$, an
arbitrary large sequence of levels will obey the Wigner-Dyson
statistics of the GUE. The ``critical'' case corresponds to
$\mu=cN^2$, with $c={\rm const}$. The situation resembles the one
which occurs in disordered electronic systems where, in the
thermodynamic limit, three distinct types of statistics corresponding
to insulator, metal and a mobility-edge system \cite{Shk93,Aro94}
exist.

In the present paper we shall take a closer look at the eigenvalue
statistics, with an emphasis on the ``critical'' case $\mu=cN^2$.  We
will show that a rather complete picture emerges already from
diagrammatic perturbation calculations, along the lines of
Refs.~\cite{Ver84,Alt95}.  In this case the two-point correlation
function $R(s)$ (smoothed out over few level spacings) differs
substantially from both, the GUE and the Poisson correlation
functions.  Here $s$ denotes the energy difference in units of the
average level spacing.  For small $c$, $R(s)$ is approximately given
by its Wigner-Dyson value, $-1/(2\pi^2s^2)$, as long as $s\ll
1/\sqrt{c}$ \cite{err}. For larger $s$, however, $R(s)$ changes sign,
reaches a maximum and eventually decreases as $1/(\pi cs^2)$.

\section{Diagrammatic Analysis}\label{diag}
To begin with, let us introduce a definition of the Rosenzweig-Porter
(RP) model which is equivalent to Eq.~(\ref{1.1}) but more convenient
for diagrammatic computation. We define
\begin{equation}\label{2.1}
H=H_0+V \;\; , \; (H_0)_{ij}= \varepsilon_i \delta_{ij} \;\; , \; V_{ii}=0
\end{equation}
where $\varepsilon_i$ are independent real random numbers with
Gaussian distribution
\begin{equation}\label{2.2}
p(\varepsilon)= \pi^{-1/2} e^{-\varepsilon^2}\, .
\end{equation}
The matrix elements $V_{ij}$ of the hermitean matrix $V$ are
independent complex random numbers with Gaussian distributed real and
imaginary parts. The distribution is determined by
\begin{equation}\label{2.3}
\BRA V_{ij}\KET=0 \; ,\;\; \BRA |V_{ij}|^2\KET = \frac{1}{2(1+\mu)}\, .
\end{equation}
It is easy to see that the probability density function of the thus
defined Hamiltonian is just Eq.~(\ref{1.1}).

The diagrammatic analysis amounts to a locator expansion of the full
single particle Green's function
\begin{equation}\label{2.5}
     {G}^{\pm} =(E^{\pm}-H)^{-1}\, ,\;\; E^{\pm} = E \pm i \delta\, \;\;
     \delta\searrow 0 \, ,
\end{equation}
with respect to the off-diagonal $V$. The unperturbed (bare) Green's
function $G_0=G\Big|_{V=0}$ is called the locator.  We consider the
density of states $\rho(E)=\tr \delta (E-H)$, its average value
\begin{equation}\label{2.8}
        \nu(E) = \BRA\BRA \rho(E) \KET_{V}\KET_{\ep} 
\end{equation}
and its correlation function
\begin{equation}\label{2.9}
        R(E,E') = \BRA\BRA \rho(E) \rho(E') \KET_{V}\KET_{\ep}
        -\nu(E)\nu(E')\, .
\end{equation}
Here $\BRA \ldots \KET_V$ ($\BRA \ldots \KET_{\ep}$) stands for
averaging with respect to the off-diagonal elements $V_{ij}$ (the
diagonal elements $\ep_i$).

We concentrate on energy separations $\omega=E'-E$ for which both
energies are close to the middle of the band, i.e.  close to $E=0$
where $\nu(E)$ is maximal. In this region, $\nu(E)$ is approximately
constant, $\nu(E)=\nu(E')$ up to ${\cal O} (1/N)$ relative
corrections. The density-density correlator $ R(s)\equiv R(E,E')/\nu^2
$ will then be a function of the dimensionless level separation
$s=\omega/\Delta$ where the average level spacing is
$\Delta=1/\nu(E=0)$.

We next analyze the spectral correlation function $R(s)$ in the regime
$s>1$ where perturbative methods are applicable. To begin with, we
decompose $R(s)$ according to $R(s)= R_1(s) + R_2(s)$ where
\begin{eqnarray}
&& R_1(s) = \frac{1}{\nu^2}\bigg\{
\av{\av{\rho(E+\om)}{V}\av{\rho(E)}{V}}{\ep}-\nonumber\\
&&\hspace{2.0cm}-\av{\av{\rho(E+\om)}{V}}{\ep}
\av{\av{\rho(E)}{V}}{\ep}\bigg\}\, ,\\
&&R_2(s) = \frac{1}{\nu^2}\bigg\{ \Big\langle
\av{\rho(E+\om)\rho(E)}{V}-\nonumber\\
&&\hspace{2.0cm}-\av{\rho(E+\om)}{V}\av{\rho(E)}{V}
\Big\rangle_{\ep}\bigg\}\, .
\label{Rdecom}
\end{eqnarray}
Note that the decomposition
$R=R_1+R_2$ is exact. The physical significance of the two functions
$R_{1/2}$ will be discussed below. Here we merely note that $R_1$
measures correlations remaining in the GUE-averaged density of states
whereas $R_2$ focuses on the GUE-correlations as such.\par
Representing the density of states in terms of the single particle
Green's function
\[
\rho(E)=-{\pi}^{-1}\imag \tr G^+(E),
\]
 and making
use of the fact that correlations (of any type) between products of
purely retarded/advanced Green's functions vanish for $N\to \infty$: $\langle
G^{+n}\rangle = \langle G^+ \rangle^n$, we obtain
\begin{eqnarray}
&&R_1=\frac{\Delta^2}{2\pi^2} \real
\Big\langle 
\tr \av{G^+(E+\om)}{V}\tr \av{G^-(E)}{V}
\Big\rangle_{\ep,c},\nonumber\\
&&R_2=\frac{\Delta^2}{2\pi^2} \real \av{
\Big\langle 
\tr G^+(E+\om) \tr G^-(E) \Big\rangle_{V,c}}{\ep},
\end{eqnarray}
where $\langle\dots\rangle_c$ denotes the connected average, $\langle
XY \rangle_c= \langle XY \rangle - \langle X \rangle \langle Y
\rangle$. Before turning to the actual calculation of these functions
let us make a few methodological remarks and introduce some building
blocks that will be of importance throughout. The whole approach will
be based on a perturbative expansion of the Green's functions in
powers of $V$. It is instructive to visualize the structure of the
expansion scheme graphically. To this end we introduce the notation
of Fig.~1(a)
where $i$ and $j$ represent matrix indices (which will not be
indicated explicitly unless necessary). As a first step of our
perturbative analysis (cf. the definition of the correlation function
$R_1$ above) we have to calculate the $V$-average of the Green's
function $G$. In a diagrammatic language, the expansion of the Green's
function can be visualized  as is shown in Fig.~1(b).
The subsequent diagrammatic analysis of this equation is simplified
drastically by two observations that hold to leading order in
$N^{-1}$. 
\begin{itemize}
\item[i)] Contributions with 'crossed GUE-lines' (see Fig.~1(e))
 are negligible \cite{Ver84}.

\item[ii)] Diagram segments which are separated from each other by
vertices $V_{ij}$ are statistically independent with respect to the
average over the on-site distribution functions $\rho(\ep)$.
\end{itemize}
The second statement is based on the fact that the indices $i$ and $j$
in a diagram like the one shown in Fig.~1(f) are eventually summed
over independently of each other. (All contributions where one of the
summations is constrained will be of higher order in $N^{-1}$.) On the
other hand, the variables $\ep_i$ at different sites are statistically
independent. Put together, these two facts imply that ii) holds to
leading order in $N^{-1}$.\par
The diagrammatic expression for the $V$-averaged Green's function
${\cal G}_i\delta_{ij}\equiv \av{G_{ij}}{V}$, subject to rule i), is
displayed in Fig.~1(c).  The statement ii) implies that to leading order
in $N^{-1}$, the self energy part $\Sigma^{\pm}$ (as shown in
Fig.~1(d)) can be replaced by the $\epsilon_i$ averaged one. We thus
obtain
\begin{eqnarray*}
&&{\cal G}_i^{\pm}\simeq
\frac{1}{G_{0,i}^{\pm-1}-\overline{\Sigma^{\pm}}},\\
&&\hspace{0.5cm}\overline{\Sigma^{\pm}}\equiv \av{\Sigma^{\pm}}{\ep} =
\frac{N}{2(1+\mu)}\av{{\cal G}^{\pm}}{\ep}.
\end{eqnarray*}
In order to solve this equation self consistently, we have to compute
the energy average of ${\cal G}$. Anticipating that a) the self energy
will be largely imaginary $\Sigma^{\pm} \simeq \mp i \Gamma $ and b)
$\Gamma \ll1$, we obtain
\begin{eqnarray*}
&&\av{{\cal G}^{\pm}}{\ep}=\frac{1}{\pi^{1/2}}\int d\ep 
\frac{e^{-\ep^2}}{E-\ep \pm i \Gamma}\simeq\nonumber\\
&&\hspace{1.0cm}\simeq \mp\frac{1}{\pi^{1/2}}\int d\ep 
\frac{e^{-\ep^2} i\Gamma }{(E-\ep + i \Gamma)(E-\ep - i \Gamma)}=
\mp i\sqrt{\pi}.
\end{eqnarray*}
The second equality is based on the assumption that the energy
argument $E\ll1$ is close to the middle of the band. As a consequence
the real part of the integral is of ${\cal O}(E)$ and negligible in
comparison with the imaginary part ${\cal O}(1)$. This justifies the
assumption a) above. Collecting everything so far we obtain the 
$V$-averaged Green's function
\begin{equation}
{\cal G}_i^{\pm}\simeq
\frac{1}{E-\ep_i\pm i\Gamma},\;\;\Gamma = \frac{N\pi^{1/2}}{2(1+\mu)}.
\label{GavV}
\end{equation}
(Note that assumption b) above holds for all $\mu \sim N^x, x>1$,
i.e. Eq.(\ref{GavV}) indeed represents the self consistent solution of the
Dyson equation in Fig.~1(c).) We next insert this expression into the
defining equation of the correlation function $R_1$ and obtain
\begin{eqnarray*}
&&R_1=\frac{\Delta^2 N}{2\pi^2} \real \frac{1}{\pi^{1/2}}
\int d\ep \frac{e^{-\ep^2}}{E+\omega-\ep+ i\Gamma}\frac{1}{E-\ep-
i\Gamma}\simeq\nonumber\\
&&\hspace{1.0cm}\simeq\frac{\Delta ^2N}{\pi^{3/2}} 
\frac{2\Gamma}{\om^2+4\Gamma^2}.
\end{eqnarray*}
Noting that the level spacing $\Delta = \pi^{1/2}/N$, we arrive at the
final result 
\begin{equation}
R_1(s)=\frac{1}{\pi c}\frac{1}{s^2+c^{-2}},
\;\;c=\frac{\Delta}{2\Gamma
}\, ,
\label{R1}
\end{equation}
for the first of the above introduced correlation functions.  We
postpone the discussion of this equation until the complementary
correlation function $R_2$ has been calculated. In principle one might
compute $R_2$ via a straightforward perturbative expansion of the
Green's function $G$. However, experience gained from the analysis of
similar correlation functions \cite{Alt95} has shown that it is
advantageous to represent the Green's functions according to
\[
G^{\pm}(E)=\partial_E \ln (E^{\pm}-H)
\]
prior to the perturbative expansion. In this way we are led to
consider
\begin{eqnarray*}
&&R_2= \real\frac{\Delta^2}{2\pi^2} \partial^2_{E',E}\Big|_{E'=E+\om}\\
&&\hspace{1.0cm} \av{ \Big\langle \tr \ln  (E'^{+}-H)\; \tr\ln(E^{-}-H) 
\Big\rangle_{V,c}}{\ep}.
\end{eqnarray*}
Expanding the logarithms in powers of $V$ and applying the
non-crossing rule i) we obtain
\begin{equation}
R_2=\frac{\Delta^2}{2\pi^2} \partial^2_{E',E}\Big|_{E'=E+\om}\real
\sum_{n=2}\frac{1}{n}\av{S_n(E,E')}{\ep} ,  \label{SN}
\end{equation}
where the diagrammatic representation of
$S_n(E,E')$ is shown in Fig.~2.
There the segments on the outer (inner) ring correspond to the Green's
function ${\cal G}^+(E')$ (${\cal G}^-(E)$) and the two rings are
connected by $n$ $V$-lines. (Note that a $n=1$ contribution is absent
because the potential $V$ is off-diagonal in the site indices.) The
rule ii) implies that each segment of the 'wheel' above can be
averaged individually over the on-site energies $\ep_i$. As a result,
the diagram $S_n$ factorizes, $S_n=\gamma^n$, $\gamma=N/(2(1+\mu))
\av{{\cal G}^+(E') {\cal G}^-(E)}{\ep}$ and $\sum_{n=2} n^{-1}
\av{S_n} {\ep}=-\ln (1-\gamma) - \gamma$. Computing the energy average
(cf. the computation of the correlation function $R_1$ above)
\[
\av{{\cal G}^+(E'){\cal G}^-(E)}{\ep}=\frac{2\pi^{1/2}i}{E'-E+2i\Gamma},
\]
and collecting constants we obtain
\begin{equation}
R_2(s)=\frac{1}{2\pi^2}\partial^2_{s}\real \left[
\ln\left(\frac{s}{s-ic^{-1}}\right)-\frac{ic^{-1}}{s-ic^{-1}} \right].
\label{2.10}
\end{equation}
We finally carry out the differentiation and add the $R_1$
contribution (\ref{R1}) to arrive at the final result
\begin{eqnarray} 
&&R(s)=\frac{1}{c\pi}\frac{1}{s^2 +  c^{-2}}+
\nonumber \\
&& + \frac{1}{2\pi^2}\LB \frac{-1}{s^2} + \frac{s^2-3c^{-2}}{(s^2
        +  c^{-2})^2} + \frac{8s^2 c^{-2}}{(s^2+  c^{-2})^3}\RB .
\label{2.10a}
\end{eqnarray}
Eq.~(\ref{2.10a}) is applicable when the energy $s\gg 1$ and fine
structures on scales $s\approx 1$ are inessential. Let us conclude this
section with a brief discussion of this result.

The contribution $R_1$
(cf.~Eq.~(\ref{R1})) has the following interpretation: The $V$-averaged
Green's function ${\cal G}$ is similar to $G_0$ except for the fact
that a finite width $\Gamma$ has been attached to each of the
uncorrelated levels $\ep_i$. This 'smearing' implies that the
corresponding correlation function $R_1$ is Lorentzian, i.e. it is a
broadened version of the $\delta$-function that would be obtained for
sharply defined autocorrelated levels.  The complementary term
$R_2$ describes correlations between the $V$ degrees of freedom. 
After combining the two contributions three qualitatively different
regimes can be identified:
\begin{itemize}
\item For $s \gg c^{-1}$ the dominant contribution comes from $R_1$
  and we obtain $R(s)\approx (\pi cs^2)^{-1}$. 
\item $c^{-1/2}\ll s \ll c^{-1}$: Still $R_1$ dominates but now $R_1
  \approx c/\pi$.  
\item $s\ll c^{-1/2}$: The $R_2$ contribution becomes the dominant one
  and we obtain the GUE-result $R_2\approx -(2\pi^2s^2)^{-1}$
  corrected by a small term $R_1\approx c/\pi$.
\end{itemize}
In summary, Eq.~(\ref{2.10a}) essentially represents a superposition of
a GUE correlation function and a smeared Poissonian auto-correlation
function.
\section{Non-Perturbative Results}\label{secnon}
The diagrammatic treatment is incapable of describing structures on
the energy scale of ${\cal O}(\Delta)$. For $c\gg 1$ an alternative
perturbation technique, applicable over the whole energy axis, can be
used. Within this approach spectral correlations are described in
terms of stochastic evolution equations~\cite{Ley90} (see also
Ref. \cite{Fre88}). In this way one obtains a spectral correlation
function $R(s)$ that depends only on the combination $s\sqrt{c}$. For
large energies, $ s\gg 1/\sqrt{c}$, the result coincides with ours,
i.e. $R(s)\approx (\pi cs^2)^{-1}$, and for small energies, $ s\ll
1/\sqrt{c}$ level repulsion sets in, i.e.  $R(s)+1 \propto cs^2$.

The complementary regime of $c\ll 1$ can be treated by Efetov's
non-perturbative supersymmetry technique \cite{Efe83} (for a recent
review see ~\cite{Fyo96}), where averages of Green's functions are
obtained from a generating functional.  The generating functional
corresponding to the RP model is similar to the one described in
\cite{FyoMir95,Fra95}. In these works the problem of a random banded
matrix with additional diagonal disorder was addressed. Taking the
band width equal to the matrix size $N$ leads to the RP model.  From
the generating functional one can obtain the correlation function of
retarded and advanced Green's functions $K^{12}=\BRA \tr
G^{+}(E+\omega)\tr G^{-}(E)\KET$ where the average is taken with
respect to Eq.~(\ref{1.1}). The final integrations can be carried out
within a saddle-point expansion the validity of which is controlled by
$N^2/\mu \gg 1$ or equivalently by $c\ll 1$, and by $\omega \ll N/\mu$
or equivalently by $s\ll c^{-1}$.  In the present work we skip the
technical details and concentrate on the discussion of the results.

In the limit $N\to \infty$ the function $K^{12}$ is given by
\begin{eqnarray}\label{3.23}
        && K^{12}(E,\omega) =  \BRA \tr G^{+} (E)\KET \BRA  \tr
G^{-}(E)\KET
 + \nonumber\\
&& \LK 1+ \frac{c}{\pi}\RK \frac{-2i}{s^2\Delta^2} 
        e^{-i\pi s}
        \sin\LK \pi s\RK + \frac{2ic}{\Delta^2 s} + \frac{2\pi c}{\Delta^2}
\, . 
\end{eqnarray}
The first term is the entirely disconnected part and the terms
of ${\cal O}(c)$ describe deviations from a pure GUE behavior. These
terms represent the analogue of the contribution $R_1$ that
appeared in the diagrammatic analysis. They result from the
correlation between the on-site energies $\varepsilon_i$. (Note that
in principle correlations of this type exist in the pure GUE as well.
In that case, however, they represent negligible ${\cal
  O}(1/N)$-effect.) From (\ref{3.23}) we
obtain the correlation function
\begin{equation}\label{3.24}
        R(s)= \LK 1+ \frac{c}{\pi}\RK 
        \LB -\LK \frac{\sin \LK \pi s\RK }{\pi s }\RK^2 + \delta(s)\RB
 + \frac{c}{\pi}\, 
\end{equation}
describing the spectral behavior in the regime $s\ll c^{-1}$, $c\ll
1$.  We next turn to the discussion of this result. We first observe
that the term $c/\pi$ equals the leading contribution of the smeared
auto-correlation $R_1(s)$ for $s\ll c^{-1}$.  For very small level
separations $R(s)$ behaves as $-1 + (1+c/\pi)(\pi s)^2/3$, i.e. apart
from a slightly modified prefactor we obtain generic GUE-behavior.
For larger values $1\ll s\ll c^{-1}$ the leading terms are identical
with those obtained in the diagrammatic treatment, as expected. (By
'leading' we mean the first order terms of an expansion in the
parameter $1/s\ll1$ after the oscillatory structure in (\ref{3.24})
has been averaged out.) In particular, the GUE behavior is only valid
up to $s\ll 1/\sqrt{c}$.  Thus, the non-perturbative results underline
the conclusion drawn from the diagrammatic analysis: $R(s)$ is
essentially a superposition of a GUE correlation with a smeared
Poissonian auto-correlation.  A conclusion to be drawn from this
observation is that the analogy between the Wigner-Dyson/Poisson
transition in the RP model and disordered electron systems,
respectively, is not complete.  In the latter case the critical
correlation function can hardly be interpreted as a simple
superposition of two terms.  This qualitative difference manifests
e.g. in the behavior of the level compressibility, $ \chi = \lim_{S\to
\infty} \int\limits_{-S}^{S} ds\, R(s) $ (where it is essential
that the limit $N\to \infty$ is taken first). The two extremes GUE
(Poisson) correspond to values $\chi=0$ ($\chi=1$). In the case of a
disordered metal at criticality the compressibility takes an
intermediate value $0< \chi < 1$ \cite{Cha96}. In the critical RP
model, however, $\chi=1$, i.e. perturbing a Poisson ensemble by a GUE
ensemble does not change the level compressibility \cite{Ley90}.
 
Finally, we would like to comment on the analysis~\cite{Guh96,Guh97}.
In these references, the RP model was solved {\it exactly} for
arbitrary values of the parameters $\mu$ and $s$. As a result of a
sophisticated combination of supersymmetry and group-theoretical
concepts Guhr \cite{Guh96} obtained non-trivial double integral
representations for the correlation functions which turned out to be
difficult to evaluate. In order to derive closed expressions for
$R(s)$ the integral was analyzed in the two limiting cases $c\gg 1$
\cite{Guh96} and $c\ll 1$ \cite{Guh97} by means of asymptotic
expansion schemes. The price to pay for the mathematical rigor of
Guhr's approach is that the physical origin of the various ingredients
to $R(s)$ is hard to identify. For this reason we believe, that a more
conventional analysis like the one discussed above was calling for.

\section{Conclusions}
We have studied the density of states correlation function $R(s)$ ($s$
measures the energy difference in units of the average level spacing)
of the Rosenzweig-Porter model.  This random matrix model contains a
parameter $\mu$ which allows to interpolate between GUE ($\mu=0$) and
Poisson statistics ($\mu=\infty$). In the thermodynamic limit $N\to
\infty$ the model shows three different types of universal functions
$R(s)$ depending on how $\mu$ scales with $N$.  From a diagrammatic
analysis (locator expansion) assisted by non-perturbative methods we
draw the following conclusions: parameter values scaling like
$\mu(N)/N^2 \to 0$ ($\mu(N)/N^2 \to \infty$) lead to GUE (Poisson)
statistics.  In the borderline case $\mu(N)/N^2\equiv c$, however, a
novel universal type of spectral behavior is observed. The
corresponding correlation function $R(s)$ has the following features:
Like in the GUE-case levels repel each other, i.e.  $R(s)\to -1$ for
$s\to 0$.  At some $c$-dependent value $s_0$, $R(s)$ changes sign,
then reaches a maximum and decreases as $(\pi cs^2)^{-1}$ for large
$s$. For $c\ll 1$ we find that the spectrum shows GUE-type statistics
up to values $s\sim 1/\sqrt{c}$. For larger values of $s$, a different
type of statistics is observed, this being a consequence of the non
GUE-correlation of the diagonal matrix elements in the RP model. These
large energy correlations can be interpreted as the tails of a widely
'smeared' energy autocorrelation of Poissonian type. We thus conclude
that the RP model in the critical case $\mu(N)/N^2\equiv c$
essentially leads to a linear superposition of Wigner-Dyson and
Poissonian behavior.
Let us finally comment on the aspect of symmetries. In this paper
we have considered the Rosenzweig--Porter model in its unitary
version. It is a straightforward matter to extend both the
diagrammatic and the non--perturbative analysis to the case of
orthogonal respectively symplectic symmetry. On the other hand,
none of our main conclusions on the structure of the models
eigenvalue statistics did depend in a {\it conceptual} way on
symmetry aspects. We thus expect the level statistics of the
models of higher symmetry to be qualitatively similar but did not
embark on any kind of quantitative analysis.

\bigskip
\bigskip

\section*{Acknowledgments}

We acknowledge the collaboration of M. Kreynin at the initial stage of
this work, as well as useful discussions with O. Prus, J.-L. Pichard
and D.L. Shepelyansky. 
We are grateful to T. Guhr for providing us with a copy of \cite{Guh97} 
prior to publication.
The research was supported by the
Israel Science Foundation administered by the Israel Academy of
Sciences and Humanities. M.J. acknowledges the support by the MINERVA
Foundation and is grateful for the support by the Institute of
Theoretical Physics at the Technion.

\bigskip
\bigskip

\noindent
Figure 1: (a) Graphical representation of the Green's function $G_0$, the
perturbation matrix elements $V_{ij}$ and its correlations.  (b),(c) The
Dyson equation for the full and the averaged Green's function, respectively.
(d) Graphical
representation of the self-energy. (e) 
A diagram with crossed GUE lines. (f) Explanation see text.

\bigskip

\noindent
Figure 2:
Graphical representation of $S_n(E,E')$  appearing  in  Eq.~(\ref{SN}).
\end{multicols}


\begin{references}
\bibitem{err} It was erroneously stated in Ref.~\cite{Sha96} that the
Wigner-Dyson statistics is applicable up to $s\ll ~ 1/c$.
\end{references}
\end{document}